\documentclass[aps,prl,longbibliography,twocolumn,superscriptaddress,nofootinbib]{revtex4-2}
\usepackage[colorlinks, urlcolor=blue,linkcolor=blue, anchorcolor=blue, citecolor=blue]{hyperref}
\usepackage[english]{babel}
\usepackage[utf8]{inputenc}
\usepackage{multirow}
\usepackage{array}
\usepackage{amsthm}
\usepackage{mathtools}
\usepackage{physics}
\usepackage{xcolor}
\usepackage{graphicx}
\usepackage{bm}
\usepackage{soul}
\usepackage{adjustbox}
\usepackage{placeins}
\usepackage[T1]{fontenc}
\usepackage{lipsum}
\usepackage{csquotes}
\usepackage{float}
\usepackage{booktabs}
\usepackage{microtype}

\usepackage[normalem]{ulem} 

\usepackage{comment}
\usepackage{lineno}
\usepackage{url}

\graphicspath{{figs/}}

\usepackage{xcolor}
\definecolor{myOrange}{rgb}{1,0.5,0.}
\definecolor{myGreen}{rgb}{0.0,0.6,0.1}

\newcommand{\removetext}[1]{} 

\newcommand{\GeVc}         {GeV/$c$}

\newcommand{\PbPb}         {\mbox{Pb--Pb}}

\newcommand{\AuAu}         {\mbox{Au--Au}}

\newcommand{\sqrtsNN}{\ensuremath{\sqrt{s_\mathrm{NN}}}}

\newcommand{\pT}           {\ensuremath{p_{\rm T}}}
\newcommand{\pTone}           {\ensuremath{p_{\rm T1}}}
\newcommand{\pTtwo}           {\ensuremath{p_{\rm T2}}}

\newcommand{\qsqr}{\ensuremath{Q^2}}

\newcommand{\vzeropT}{\ensuremath{v_{0}}(\pT)}

\newcommand{\npT}{\ensuremath{n(\pT)}}

\newcommand{\VzeroMatrix}{\ensuremath{V_{0}}}
\newcommand{\argpTonepTtwo}{\ensuremath{(p_{\rm T1},p_{\rm T2})}}

\newcommand{\lambdaone}{\ensuremath{\lambda_1}}
\newcommand{\lambdatwo}{\ensuremath{\lambda_2}}
\newcommand{\lambdaratio}{\ensuremath{\lambdatwo/\lambdaone}}

\newcommand{\app}{Supplemental Material} 

\begin{document}

\title{Dynamical correlations across momentum scales in the Quark--Gluon Plasma}
\author{L.~Du}
\affiliation{Nuclear Science Division, Lawrence Berkeley National Laboratory, Berkeley CA 94720}
\affiliation{Department of Physics, University of California, Berkeley CA 94720}
\author{P.~M.~Jacobs}
\affiliation{Nuclear Science Division, Lawrence Berkeley National Laboratory, Berkeley CA 94720}
\date{\today}
\affiliation{Department of Physics, University of California, Berkeley CA 94720}


\begin{abstract}

Experimental probes of the Quark-Gluon Plasma (QGP) generated in heavy--ion collisions span a broad range in momentum scale: low transverse momentum (low \pT) measurements probe collective dynamics, while high \pT\ measurements probe the response to QGP excitation by jets (jet quenching). However, the dynamical interplay between QGP collective dynamics and jet quenching is currently poorly understood. We present a new framework for exploring dynamical correlations across momentum scales in heavy--ion collisions, based on the \pT--differential radial-flow observable \vzeropT. Measured \vzeropT\ phenomenology is traced to the evolution in strength and coherence of distinct underlying fluctuation modes. We then propose new experimental observables to quantify this evolution. The eigenvalue spectrum of the reference-aligned covariance matrix \VzeroMatrix\ characterizes the effective fluctuation rank, while the ratio \lambdaratio\ provides a compact measure of the leading departure from single-mode behavior. The \pT-dependence of the corresponding eigenvectors then maps the evolution from a single coherent soft mode to multi-mode dynamics including coalescence and jet quenching. These novel, experimentally accessible observables provide unique insight into dynamical correlations across momentum scales in the QGP.
\end{abstract}


\maketitle


\paragraph{\bf Introduction.}

Strongly-interacting matter at very high temperature and density forms a Quark--Gluon Plasma (QGP), in which quarks and gluons are deconfined~\cite{Shuryak:2014zxa,Busza:2018rrf}. A QGP filled the early universe, and QGP is generated and studied today by the interaction of heavy nuclei at the Relativistic Heavy Ion Collider (RHIC) and the Large Hadron Collider (LHC)~\cite{Busza:2018rrf,Harris:2023tti}. Experimental measurements at these facilities, together with theoretical calculations, show that the dynamical evolution of the QGP exhibits coherent collective flow~\cite{Heinz:2013th,Gale:2013da,Huovinen:2006jp}, and that the QGP interacts with hard jets and responds to the corresponding excitation (``jet quenching'')~\cite{Cunqueiro:2021wls,Apolinario:2022vzg,Wang:2025lct}.

These measurements span a broad range in momentum scale. Collective behavior, which is characterized by correlation over a large phase space interval, is probed by particles with ``soft'' transverse momentum, $\pT\!\lesssim\!3$~\GeVc. In contrast, jets are generated in high momentum transfer (high \qsqr) partonic interactions; jet quenching phenomena are typically characterized by particles with ``hard'' $\pT\gtrsim10$ \GeVc, and by measurements of reconstructed jets at high momentum scales. The intermediate \pT\ region exhibits phenomena characteristic of a mixture of soft and hard processes, notably quark coalescence~\cite{Fries:2003vb,Greco:2003xt,Molnar:2003ff,Fries:2008hs,Becattini:2009fv}. 

Collective flow and jet quenching are usually treated theoretically as distinct phenomena, to simplify their modeling in comparison to data (e.g.~\cite{JET:2013cls,JETSCAPE:2020shq,JETSCAPE:2020mzn,Nijs:2020roc, Nijs:2020ors,Casalderrey-Solana:2018wrw,JETSCAPE:2024cqe}). Soft anisotropic flow coefficients constrain the bulk geometry and transport  properties~\cite{Romatschke:2007mq,Heinz:2009xj,Song:2010mg}, and 
coalescence studies characterize quark recombination in the 
intermediate-\pT\ region~\cite{Fries:2008hs,Becattini:2009fv}. Jet quenching probes the microscopic and coherence structure of the QGP~\cite{Cunqueiro:2021wls,Apolinario:2022vzg,Wang:2025lct} and its response to excitation~\cite{CMS:2011iwn,ALICE:2023qve,STAR:2025yhg}, and constrain jet transport properties~\cite{JET:2013cls,Apolinario:2022vzg,JETSCAPE:2024cqe}.

However, QGP phenomena in these different momentum regimes should be coupled physically; for instance jet quenching is strongly influenced by long--range coherence effects that are the QCD analogue to the Landau--Pomeranchuk--Migdal effect in QED~\cite{Wang:2025lct}, and the QGP response to jet excitation should have a collective component~\cite{Cao:2022odi}.  An understanding of the dynamical correlation of QGP phenomena at different momentum scales will help elucidate the nature and limits of hydrodynamic collectivity, the mechanisms underlying coalescence, and the constraints on QGP structure and dynamics from jet quenching measurements.

Study of such dynamical correlations requires the measurement of event-by-event fluctuations by an observable that links different momentum scales, quantifying the degree to which the responses at different scales originate from a single collective degree of freedom or from multiple, partially decorrelated dynamical modes. A sensitive and experimentally accessible probe of such correlations is the \pT--differential radial-flow observable \vzeropT~\cite{Gardim:2019iah,Schenke:2020uqq},
%
\begin{equation}
\vzeropT
=\frac{\langle \delta n(\pT)\,\delta[\pT]_{\rm ref} \rangle}
       {\langle n(\pT)\rangle\,\sigma_{[\pT]_{\rm ref}}} ,
\label{eq:v0_def}
\end{equation}
%
\noindent
where $n(\pT)$ is the event-wise normalized particle yield in a narrow bin at momentum $\pT$,
$\delta n(\pT)\!=\!n(\pT)\!-\!\langle n(\pT)\rangle$, and $\langle\cdots\rangle$ is the event--ensemble average. The label ``ref'' indicates a disjoint reference particle sample with event-wise mean transverse momentum $[\pT]_{\rm ref}$, and fluctuations
$\delta[\pT]_{\rm ref}\!=\![\pT]_{\rm ref}\!-\!\langle[\pT]_{\rm ref}\rangle$ with
variance $\sigma_{[\pT]_{\rm ref}}$ relative to the ensemble--averaged distribution. 
Thus, the numerator measures the covariance between the yield at $\pT$ and a
global momentum scale, while the denominator normalizes this correlation
by the average spectrum and the fluctuation amplitude of the reference
observable.
In experiments, the reference quantities and particle distribution of interest are measured in separated pseudorapidity intervals, to suppress non-flow effects~\cite{ATLAS:2025ztg,ALICE:2025iud}.

\begin{figure*}[htbp]
  \centering
  \includegraphics[width=0.75\linewidth]{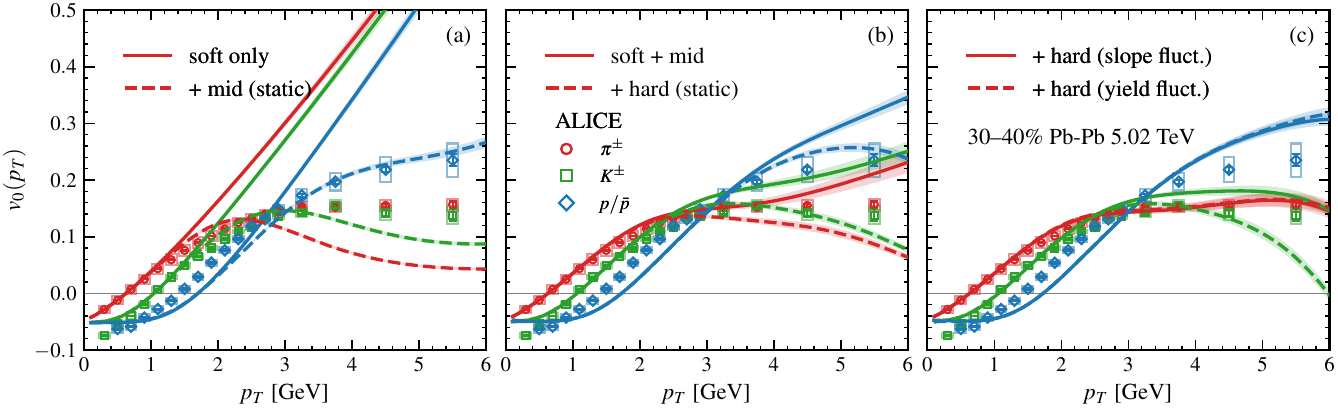}
    \caption{
    Three-stage construction of \vzeropT\ for $\pi^{\pm}$, $K^{\pm}$, and
    $p/\bar p$ in 30--40\% \PbPb\ collisions at $\sqrtsNN=5.02$~TeV.
    (a) Soft-only fluctuations (solid) and soft fluctuations with a static
    mid-\pT\ component (dashed).
    (b) Soft+mid fluctuations (solid), augmented by a
    static hard tail (dashed).
    (c) Soft+mid+hard fluctuations with slope modulation (solid) or yield
    modulation (dashed) of the hard component.
     Shaded bands denote statistical
    uncertainties from bootstrap resampling. ALICE data from~\cite{ALICE:2025iud}. 
    }
  \label{fig:v0_buildup}
\end{figure*}

Recent LHC measurements of \vzeropT\ reveal a rise--plateau--downturn structure, with species hierarchy and robust low-\pT\ universality~\cite{ATLAS:2025ztg,ALICE:2025iud}.  
While these features encode the interplay of soft collectivity, quark 
recombination, and jet quenching, their dynamical origin remains 
unclear~\cite{Parida:2024ckk,Jahan:2025cbp,Du:2025dpu,Jia:2025rab,Bhatta:2025oyp,Saha:2025nyu,Sambataro:2025pop,Wan:2025rzg}. In particular, existing soft-sector calculations generically predict a monotonic rise of \vzeropT\ with \pT, indicating that the experimentally observed plateau and downturn point to additional sources of fluctuation and decorrelation beyond a single collective mode.

In this Letter we propose a novel approach to quantifying dynamical correlations of particle production across momentum scales in heavy-ion collisions, utilizing a covariance matrix which is based on the \vzeropT\ observable. We first analyze current \vzeropT\ measurements in terms of underlying fluctuation modes, showing that the observed structure arises from a sequence of coherent, partially coherent, and anti-correlated responses of soft, coalescence, and hard components. We then introduce the reference-aligned covariance matrix \VzeroMatrix\argpTonepTtwo\ and show that its eigenvalue spectrum characterizes the effective fluctuation rank, while the eigenvalue ratio \lambdaratio\ provides a compact diagnostic of soft--hard decorrelation and the emergence of subleading fluctuation structure, revealing information not accessible from the projected observable \vzeropT\ alone. This framework provides a direct and experimentally accessible means to map the dynamical correlation of soft, mid-, and hard momentum sectors in heavy--ion collisions, providing new insight into QGP dynamics and its collective--to--partonic transition.


\paragraph{\bf Spectrum composition and correlation topology.}
We focus in this analysis on the inclusive particle spectrum in the range $\pT\!<\!15$ \GeVc. 
We utilize a parametrized Multi-component Model to provide qualitative insight into the interplay of dynamical correlations across momentum scales. The inclusive particle spectrum is modeled as the sum of three overlapping components, each dominant in 
a different \pT\ regime (see \app~\cite{suppl}):
%
\begin{equation}
n(\pT)=n_{\rm soft}(\pT)+n_{\rm mid}(\pT)+n_{\rm hard}(\pT).
\end{equation}
%
\noindent
The soft--sector distribution ($\pT\!\lesssim\!2$~\GeVc) has a blast-wave form, with parameters corresponding to freeze-out temperature and radial flow~\cite{Schnedermann:1993ws,ALICE:2019hno}. The intermediate $\pT$ region ($2\!\lesssim\!\pT\!\lesssim\!5$~\GeVc) 
is modeled by a recombination-inspired component that remains correlated 
with the soft background, reflecting their shared hydrodynamic origin.
The hard tail ($\pT\!\gtrsim\!5$~\GeVc) has a power-law form matched continuously to the coalescence region.

Event-by-event fluctuations of the spectrum components are introduced by correlated variation in the
soft and mid-\pT\ parameters, generating partially
coherent deformations of their shapes.  
Hard-sector fluctuations have two quenching--like components: yield modulation that rescales the total yield in the hard region, and slope modulation
that varies the power-law index.  
Together, these fluctuations generate the full soft--mid--hard covariance
structure that governs \vzeropT\ and \VzeroMatrix\argpTonepTtwo.

We stress that the purpose of this Multi-component Model is to provide qualitative insight into cross-scale dynamical correlations. It is not intended as a precision description of existing measurements, nor has it been tuned for quantitative agreement.

This sequential construction enables a controlled analysis of how the
soft, mid-, and hard-momentum sectors contribute to the inclusive spectrum
$\npT$ and to \vzeropT, while distinguishing normalization-induced effects
from genuine dynamical (de)correlations.
To make this explicit, the spectrum may be written schematically as
$\npT = n_{\rm soft}(\pT) + n_{\rm add}(\pT)$,
where $n_{\rm add}$ denotes additional mid- or hard-momentum contributions.
From the definition of \vzeropT\ in Eq.~\eqref{eq:v0_def}, the numerator measures
the covariance between the yield at $\pT$ and the reference momentum scale,
while the denominator is proportional to the mean yield $\langle n(\pT)\rangle$.
If an added component contributes to the mean spectrum but is uncorrelated
with the reference, it increases the denominator without contributing to the
numerator, leading to a suppression of \vzeropT\ purely through normalization.
This dilution effect does not reflect a loss of correlated dynamics, but rather the
superposition of statistically independent sources.
By contrast, an added component with coherent or anti-coherent fluctuations
enhances or suppresses the signal dynamically (see \app{}~\cite{suppl}).

Figure~\ref{fig:v0_buildup} shows a buildup of \vzeropT\ through 
sequential activation of the soft, mid-, and hard components, 
separately for pions, kaons, and protons, compared to ALICE data~\cite{ALICE:2025iud}. The event-wise average $[\pT]_{\rm ref}$ is calculated using the pion spectrum as a proxy for the charged-hadron
distribution. Panel (a) shows only the soft (blast-wave) sector fluctuations (solid). Both the monotonic increase in \vzeropT\ with \pT\ and the particle species ordering are consistent with the data for $\pT\!\lesssim\!1.5$~\GeVc\ and with hydrodynamic calculations \cite{Schenke:2020uqq,Jahan:2025cbp,Parida:2024ckk,Du:2025dpu}, corresponding to the baseline soft-sector interpretation commonly adopted in the literature. A static mid-\pT\ component (coalescence-like) increases the denominator of Eq.~\eqref{eq:v0_def} in the $2<\pT<5$~\GeVc\ range, thereby flattening \vzeropT\ (dashed); this is a 
dilution rather than a genuine dynamical decorrelation. 

Figure~\ref{fig:v0_buildup}, panel (b), introduces fluctuations in the  mid-\pT\  component, which partially restores the covariance. 
However, since these fluctuations retain incomplete coherence with the soft background, 
they generate the plateau over $2\!<\!\pT\!<\!4$~\GeVc\ (solid), in contrast to the 
linear rise produced by a fully coherent soft sector.
The observed baryon--meson ordering at mid-\pT\ follows from recombination, since baryons inherit flow from a larger number of constituent quarks, generating stronger soft--mid correlations. A static hard tail (dashed) suppresses \vzeropT. Because the spectrum $n(\pT)$ is
event-wise normalized, the static hard component acquires a small 
correlation with the reference through overall multiplicity fluctuations.
This effect is purely normalization-induced and likewise does not constitute a genuine
soft--hard dynamical coupling.

Finally, Fig.~\ref{fig:v0_buildup}, panel (c), introduces event--wise fluctuations in the yield normalization and power--law index (slope) of the hard component. Both effects can arise from jet quenching but generate distinct effects in \vzeropT: slope fluctuations produce a largely uniform flattening, while yield fluctuations produce a marked turnover for kaons. The component ratios (\app) show that  soft, coalescence, and hard contributions make similar contributions to the kaon spectrum for $3\!<\!\pT\!<\!5$~\GeVc. In this region the soft--hard anti-correlation due to
yield modulation acts in opposition to the positive soft--mid
covariance, thereby cancelling the coherent response and suppressing \vzeropT\ most efficiently for kaons. For pions, the hard tail is dominant only at larger \pT\ and the cancellation is weaker. The proton spectrum is largely dominated by the soft component in this range and retains an almost coherent response. The region $3\!<\!\pT\!<\!5$~\GeVc{} for kaons thus provides discrimination of the mechanisms underlying hard-sector fluctuations: a strong downturn \vzeropT\ indicates yield
modulation, while a flatter tail corresponds to slope modulation.

Figure~\ref{fig:v0_buildup} thus maps the rise--plateau--downturn structure of
\vzeropT\ onto three distinct correlation regimes:
positive soft--soft covariance at low \pT,
partial soft--mid coherence at intermediate \pT,
and negative soft--hard covariance combined with dilution at high \pT. While \vzeropT\ captures these effects in projection, it does not by itself
disentangle normalization-induced dilution from genuine dynamical correlations,
nor does it reveal how many independent fluctuation modes are active or where
their correlations are anchored in momentum space.
This motivates the development of a novel covariance-based and eigenmode-resolved
analysis, which we introduce next.

\begin{figure}[t]
    \centering
    \includegraphics[width=0.75\linewidth]{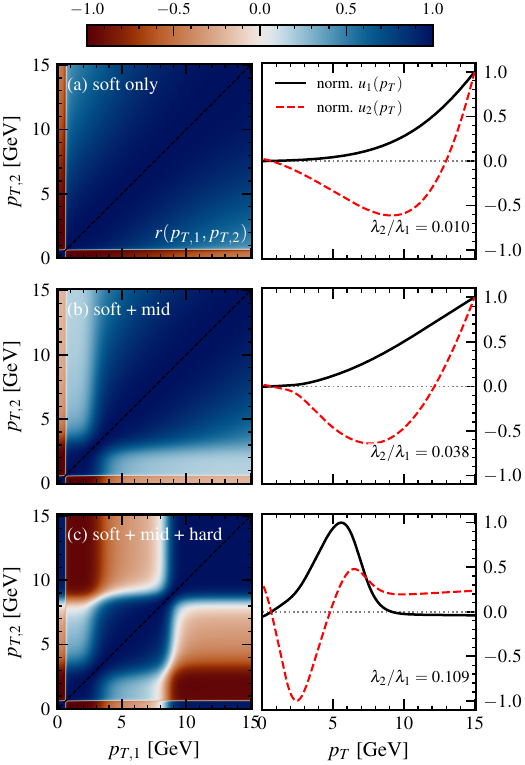}
    \caption{
    Diagnostics of the reference-aligned covariance matrix
    $V_0\argpTonepTtwo$ for pions.
    Left panels: two-bin factorization ratio $r\argpTonepTtwo$ for
    (a) soft-only,
    (b) soft+mid, and
    (c) soft+mid+hard (slope-modulated) configurations.
    Right panels: corresponding normalized leading and subleading
    eigenvectors $u_{1}(\pT)$ and $u_{2}(\pT)$, together with the
    subleading-to-leading eigenvalue ratio $\lambda_2/\lambda_1$.
    }
    \label{fig:v0_diagnostics}
\end{figure}

\paragraph{\bf Reference-aligned covariance matrix and fluctuation structure.}

We define the reference-aligned covariance matrix to measure the \pT-dependent strength and mutual correlation of different fluctuation modes,
%
\begin{equation}
\VzeroMatrix\argpTonepTtwo
=\Big\langle\big[\delta n(\pTone)\,\delta[\pT]_{\rm ref}\big]
          \big[\delta n(\pTtwo)\,\delta[\pT]_{\rm ref}\big]\Big\rangle,
\end{equation}
%
\noindent
where $\delta[\pT]_{\rm ref}$ serves as a filter to isolate the spectrum component that co-varies with the reference scale $[\pT]_{\rm ref}$. The diagonal elements $V_0(\pT,\pT)$ quantify the mean-squared amplitude of the reference-aligned fluctuations $\delta n(\pT)\delta[\pT]_{\rm ref}$, while the off-diagonal elements quantify how these fluctuations vary together between different \pT\ regions, directly probing soft--mid--hard correlations. A single fully coherent fluctuation mode corresponds to a rank--1 structure of $V_0$, while deviations from rank one indicate additional semi-independent fluctuation modes, analogous to flow-factorization breaking in anisotropic flow PCA analyses \cite{Bhalerao:2014mua,Mazeliauskas:2015efa,Gardim:2019iah}. We emphasize that $\VzeroMatrix\argpTonepTtwo$ contains independent and complementary information to \vzeropT, providing unique insight into dynamical correlations across momentum scales.

The information content of $\VzeroMatrix\argpTonepTtwo$ can be characterised by two novel observables. The first is the factorization ratio,
%
\begin{equation}
r\argpTonepTtwo
= \frac{V_0\argpTonepTtwo}
        {\sqrt{V_0(\pTone,\pTone)V_0(\pTtwo,\pTtwo)}},
\label{Eq:rFactorize}
\end{equation}
%
\noindent
which maps deviations from a rank--1, fully coherent structure.
The eigenvalue decomposition $V_0=\sum_i\lambda_i\,u_i(\pTone)u_i(\pTtwo)$
identifies the underlying fluctuation components. The leading mode $u_1(\pT)$ captures the coherent hardening of the spectrum, with additional modes $u_{2,3,\dots}(\pT)$ becoming significant when new fluctuation sources are active. The second new observable is the subleading-to-leading eigenvalue ratio \lambdaratio\footnote{Here \lambdaratio\ is used as a compact diagnostic of the leading departure from a rank--1 fluctuation structure, rather than as a unique count of the number of active modes. If higher eigenvalues $\lambda_3,\lambda_4,\ldots$ are non-negligible, the full eigenvalue spectrum $\{\lambda_i\}$ is required to characterize the effective fluctuation rank. The numerical values of \lambdaratio\ obtained in the present simplified model are therefore model dependent and should not be interpreted as universal thresholds for mode activation.}; the emergence of significant subleading fluctuation structure is reflected in a distinct $u_2(\pT)$ together with a rise in \lambdaratio. The rank structure of $V_0$ therefore quantifies the emergence, independence, and relative strength of additional dynamical fluctuation modes, beyond the information encoded in the projected observable $v_0(p_T)$.

Figure~\ref{fig:v0_diagnostics} presents these diagnostics for pions. The
factorization ratio $r(\pTone,\pTtwo)$ directly maps how event-by-event yield
fluctuations in different momentum bins are correlated, with its sign and
magnitude indicating correlated, decorrelated, or anti-correlated behavior
across momentum scales.
Panel~(a), with soft-only fluctuations, exhibits excellent
factorization ($r\simeq1$) and small $\lambdaratio\simeq10^{-2}$.
The leading eigenvector $u_1(\pT)$ rises monotonically with \pT, consistent
with a single collective hardening mode, with the subleading mode $u_2(\pT)$ exhibiting
only the sign changes required by orthogonality.

Panel~(b), including mid-\pT\ fluctuations, exhibits factorization breaking at $\pT\gtrsim3$~\GeVc\ (left), indicating decorrelation 
between soft and mid-$\pT$ sectors. Consistently, \lambdaratio\ increases fourfold to 
$\sim0.04$, demonstrating the appearance of a second 
fluctuation mode. While the overall shape of $u_2(\pT)$ continues to be dominated by the soft-mode orthogonality pattern, the subleading structure is shifted toward 
the 2--5~\GeVc\ region where coalescence fluctuations enter.
The form of the dominant mode $u_1(\pT)$ becomes slightly steeper in $2<\pT<5$~\GeVc,
reflecting the redistribution of variance from the soft mode to the semi-independent coalescence-driven mode.

Panel~(c), which also includes hard-sector fluctuations, generates a
stronger low--to--high--\pT\ anti-correlation, visible as 
the red negative band in the factorization map (left), with \lambdaratio\ increasing to $\sim0.1$. The subleading eigenvector becomes ``tilt-like,'' with opposite sign at
low and high momentum, indicating a hard-tail fluctuation superimposed on
the soft baseline. Such sign-changing cross-scale correlations and the emergence of an
independent hard-sector fluctuation mode are not accessible from
$v_0(p_T)$ alone.

\begin{figure}[t]
    \centering
    \includegraphics[width=0.65\linewidth]{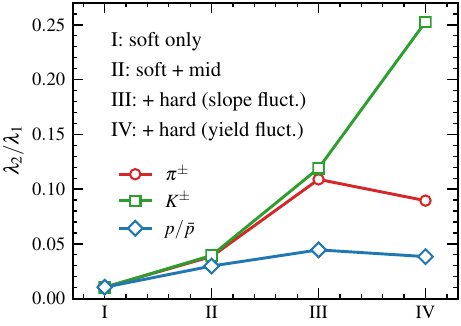}
    \caption{
    Subleading-to-leading eigenvalue ratio \lambdaratio\ for
    $\pi^\pm$, $K^\pm$, and $p/\bar p$ for different sets of fluctuation modes.
    }
    \label{fig:lambda_ratio}
\end{figure}

Figure~\ref{fig:v0_diagnostics} reveals a
controlled increase in the effective rank of $V_0$: a single coherent
soft mode, followed by more complex multi-mode dynamics due to mid- and hard-sector
variations. The value of \lambdaratio, which measures the strength of the leading subleading mode relative to the dominant mode, provides a compact diagnostic of this transition. Figure~\ref{fig:lambda_ratio} shows \lambdaratio\ for 
$\pi^\pm$, $K^\pm$, and $p/\bar p$ for these choices of fluctuation mode. For soft-only (case I), all species have $\lambdaratio\simeq0.01$, reflecting the common single-mode origin.  
Mid-\pT\ fluctuations (case~II) increase $\lambdaratio$ for all species and 
generate a baryon--meson separation, consistent with expectations from coalescence.  
Hard--sector variations (cases~III and IV) increase the ratio more strongly, 
producing a clear hierarchy 
$\lambdaratio(K)>\lambdaratio(\pi)>\lambdaratio(p)$ 
that reflects the differing degree of soft--mid--hard mixing for each species.

This kaon enhancement arises from the same soft--mid--hard mixing that shapes $v_0^K(p_T)$ (Fig.~\ref{fig:v0_buildup}(c)): the soft, coalescence, and hard contributions to the kaon spectrum are comparable in the region $3<p_T<5$ GeV, making $K^\pm$ especially sensitive to decorrelation among these sectors. The resulting rise in \lambdaratio\ demonstrates how the eigenvalue spectrum exposes the underlying fluctuation dimensionality behind the species-dependent \vzeropT\ response. Unlike \vzeropT, which provides only a projected measure of cross-scale correlations, the experimentally accessible \lambdaratio\ provides a compact measure of the departure from a single dominant fluctuation mode.


\paragraph{\bf Experimental considerations.} The quantities $V_0\argpTonepTtwo$,
$r\argpTonepTtwo$, and \lambdaratio, are experimentally measurable, using techniques similar to those used for measuring high-order flow
harmonics~\cite{Bhalerao:2014mua,Mazeliauskas:2015efa,ALICE:2011svq,
CMS:2011cqy,ALICE:2012eyl}. However, their measurement with good statistical precision is limited in range at high \pT, due to the sharply falling inclusive particle production spectrum~\cite{STAR:2003fka,CMS:2016xef,ALICE:2018vuu}. Current measurements of charged--particle  \vzeropT\ for \PbPb\ collisions at $\sqrtsNN\!=\!5.02$ TeV~\cite{ATLAS:2025ztg,ALICE:2025iud} and \AuAu\ collisions at $\sqrtsNN\!=\!200$ GeV~\cite{STARoverviewIS25} extend to $\pT\!\sim\!10$ \GeVc\ with good precision, well into the region where hard modes are expected to dominate, enabling the analysis program outlined here. 

This momentum reach covers the region of primary interest for species-resolved measurements, $2\!<\!\pT\!<\!5$~\GeVc\ region, where our framework predicts the largest mode-dependent differences, most notably for kaons. This approach may also be applicable to RHIC Beam Energy Scan data~\cite{STAR:2008med,STAR:2017sal,Bzdak:2019pkr,Du:2024wjm}, where the relative strength of soft, coalescence, and Cronin scattering~\cite{Antreasyan:1978cw} contributions is expected to evolve rapidly with \sqrtsNN.

\paragraph{\bf Summary.}

We have presented a novel framework and new observables to quantify the dynamical interplay of the mechanisms underlying particle production in the QGP: collective flow, partonic coalescence, and jet quenching. The framework is based on the previously-defined radial flow fluctuation observable \vzeropT, which correlates event-by-event fluctuations at various momentum scales with those of a reference quantity that characterizes the event structure as a whole. The new framework is explored using a simplified model of heavy--ion collisions that incorporates fluctuations in soft hydrodynamic emission, quark recombination, and quenched hard fragments. The model reproduces qualitatively the rise--plateau--downturn structure and species hierarchy of \vzeropT\ observed in LHC data, showing how coherent soft response, partially coherent coalescence, and decorrelated or anti-correlated hard modes combine across momentum scales.

The new observables characterize the covariance matrix \VzeroMatrix\argpTonepTtwo, which probes dynamical correlations across momentum scales and provides complementary information to \vzeropT. The factorization ratio $r(\pTone,\pTtwo)$ and the eigenvalue spectrum of \VzeroMatrix\argpTonepTtwo\ quantify the transition from soft coherence to multi-mode dynamics, with the subleading-to-leading eigenvalue ratio $\lambda_2/\lambda_1$ providing a compact measure of the leading departure from rank--1 behavior. This mode-based picture elucidates the systematic pattern of \vzeropT\ observed in current measurements. In particular, the observed low-$\pT$ universality of \vzeropT{}, both in shape
and in $\pT/\langle \pT\rangle$ scaling, follows directly from the dominance of a
single soft fluctuation mode \cite{single_mode}.

These results establish \VzeroMatrix\argpTonepTtwo\ as a unique, experimentally accessible probe of the dynamical interplay of collective modes of excitation, partonic coalescence, and jet quenching in the QGP, a long--standing goal in the field. Measurement of the proposed observables will quantify the correlation topology linking collective and partonic physics in the QGP. When confronted with data, these observables provide new structural constraints on theoretical descriptions, revealing the effective dimensionality and coherence pattern of dynamical fluctuations across momentum scales.

\paragraph*{\bf Acknowledgements.}

This work was supported by the US Department of Energy, Office of Science, Office of Nuclear Physics under grant number DE-AC02-05CH11231.


\bibliography{refs}

\appendix

\setcounter{figure}{0} 
\setcounter{equation}{0} 

\section*{\app}
\subsection{Decoding $v_0(p_T)$: composition, correlations, and fluctuation rank}

The analytic structure of $v_{0}(p_T)$ may be understood through three
connected layers:  
(a) the weighted composition of the spectrum,  
(b) its mapping to underlying inter-sector correlations, and  
(c) the fluctuation rank of the associated covariance matrix.  
These relations hold independent of model details and provide the
theoretical basis for the interpretation of the results in the main text.

\paragraph{Weighted composition.}

Because $v_0(p_T)$ depends on both the covariance (numerator) and the mean
yield (denominator), a flattening of $v_0(p_T)$ does not necessarily
indicate a loss of soft correlations—it may simply reflect dilution from an
additional, weakly correlated yield component.  Consider a two-component
spectrum with a fluctuating soft contribution $n_s(p_T)$ and a static or
weakly correlated component $n_x(p_T)$, such that
$n(p_T)=n_s(p_T)+n_x(p_T)$.  The measured fluctuation becomes
\begin{equation}
v_0(p_T)
=\sum_{k=s,x} f_k(p_T)\,v_0^{(k)}(p_T),
\label{eq:weighted_sum_SM}
\end{equation}
where
\begin{equation}
f_k(p_T)=
\frac{\langle n_k(p_T)\rangle}
     {\langle n_s(p_T)+n_x(p_T)\rangle},
\end{equation}
a weighted sum of the intrinsic sector responses $v_0^{(k)}(p_T)$.  
If the added source is uncorrelated with the reference
($v_0^{(x)}\!\simeq\!0$),
\begin{equation}
v_0(p_T)\simeq f_s(p_T)\,v_0^{(s)}(p_T),
\label{eq:dilution_SM}
\end{equation}
corresponding to pure dilution of the coherent soft mode.
Partially correlated or anti-correlated components modify the numerator as
well, producing precisely the rise–plateau–downturn structure observed in
experiment.  This compact formulation directly links the observed $v_0(p_T)$
to the underlying spectral composition.

\paragraph{Covariance matrix and fluctuation rank.}

The covariance matrix of normalized yield fluctuations,
\begin{equation}
V_0(p_{T1},p_{T2})
=\frac{\langle\delta n(p_{T1})\,\delta n(p_{T2})\rangle}
{\langle n(p_{T1})\rangle\,\langle n(p_{T2})\rangle},
\label{eq:V0_SM}
\end{equation}
reveals the number of independent fluctuation drivers through its
eigen-decomposition
\begin{equation}
V_0(p_{T1},p_{T2})
=\sum_i \lambda_i\,u_i(p_{T1})\,u_i(p_{T2}),
\label{eq:V0_eigen_SM}
\end{equation}
which defines orthogonal modes $u_i(p_T)$ with strengths $\lambda_i$.
A nearly factorized (rank--1) matrix corresponds to a single coherent soft
mode, while partially decorrelated mid- or hard-sector variations generate
subleading eigenvalues $\lambda_{2,3,\dots}$.
The ratio $R_{\rm rank}\equiv\lambda_2/\lambda_1$ therefore provides a
compact and quantitative diagnostic of fluctuation dimensionality.

Physically, $u_1(p_T)$ represents the global hardening deformation driven by
collective flow; a subleading $u_2(p_T)$ localized near 2--5~GeV reflects
coalescence-type fluctuations, whereas a low–high $p_T$ sign change signals
soft–hard anti-correlation.

In the main analysis we employ the \emph{reference-aligned} covariance,
constructed from the projected fluctuations
$\delta n(p_T)\,\delta[p_T]_{\rm ref}$ so that it isolates the part of the
spectrum that fluctuates coherently with the soft reference in $v_0(p_T)$.
This aligned covariance enhances coherent contributions but retains the same
eigenvector structure and rank ordering as the normalized covariance.  The factorization ratio
\begin{equation}
r(p_{T1},p_{T2})
=\frac{V_0(p_{T1},p_{T2})}
      {\sqrt{V_0(p_{T1},p_{T1})\,V_0(p_{T2},p_{T2})}},
\label{eq:r_SM}
\end{equation}
provides a complementary measure of local correlations: $r\simeq 1$
indicates a single coherent (rank--1) mode, $0<r<1$ signals partial
decorrelation, and $r<0$ indicates anti-correlation between momentum
regions.  A combined analysis of $r(p_{T1},p_{T2})$ and $R_{\rm rank}$ thus
pinpoints where factorization breaks and how many fluctuation modes
contribute.

\subsection{Spectral framework and fluctuations}

To model these mechanisms quantitatively, the transverse-momentum
distribution ($p_T<15$~GeV) is represented as a superposition of
three physically motivated sectors—a soft blast-wave baseline, an
intermediate coalescence bridge, and a hard fragmentation
tail—each corresponding to a distinct microscopic regime.  The inclusive
spectrum
\begin{equation}
\frac{dN}{p_Tdp_T}
= 
\big[
N_{\rm soft}S(p_T)
+N_{\rm mid}C(p_T)
+N_{\rm hard}H(p_T)
\big],
\end{equation}
uses normalized component shapes $S$, $C$, and $H$ with smooth 
transitions around $p_T\simeq2$ and $5$~GeV.  The soft component
follows an isotropic blast-wave form with kinetic temperature
$T_{\rm kin}$, surface velocity $\beta_s$, and profile exponent $n$,
\[
S(p_T)\!\propto\!
\int_0^1 ds\, s\, m_T
I_0\!\!\left(\frac{p_T\sinh\rho_T}{T_{\rm kin}}\right)
K_1\!\!\left(\frac{m_T\cosh\rho_T}{T_{\rm kin}}\right),
\]
where $\rho_T=\tanh^{-1}(\beta_s s^n)$ and $m_T=\sqrt{p_T^2+m^2}$.
The coalescence component employs an effective quark kernel
$f_q(p_T)\!\propto\!p_T^{a_q}\exp[-p_T/T_q^{\rm eff}]$ with
$T_q^{\rm eff}=T_{q}(\beta_{\rm eff}/\beta_{\rm ref})^{\eta}$,
leading to
\[
C_h(p_T)\!\propto\!
\left[f_q(p_T/n_q;a_q,T_{q},\eta,\beta_{\rm ref})\right]^{n_q}
\Big(1+\frac{p_T}{p_s}\Big)^{-\delta},
\]
while the high-$p_T$ tail is described by a quenched power law
\[H(p_T)\!\propto\!p_T^{-1}(1+p_T/p_0)^{-n_0}.\]

\begin{figure}
    \centering
    \includegraphics[width= 0.8\linewidth]{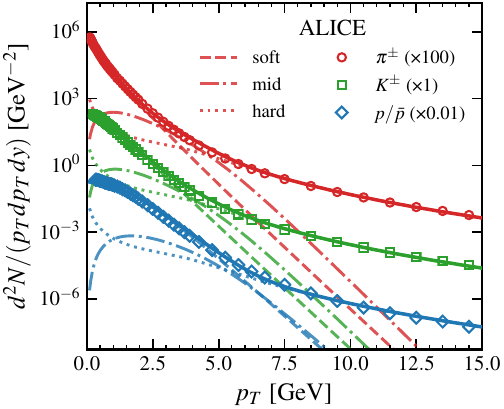}
    \caption{Inclusive transverse-momentum spectra of $\pi^{\pm}$, $K^{\pm}$, and
    $p/\bar p$ in 30--40\% Pb--Pb collisions at
    $\sqrt{s_{NN}}=5.02$~TeV \cite{ALICE:2018vuu} compared with the three-component model.
    Solid curves show the total spectra, while others indicate the
    soft, mid-$p_T$ (coalescence-like), and hard (fragmentation-like)
    components.
  }
    \label{fig:spectra}
\end{figure}

Event-by-event fluctuations are introduced by correlated variations of
$(\beta_s,\beta_{\rm mid},T_{\rm kin},T_q)$ drawn from a four-dimensional
multivariate normal distribution with covariances
$\rho_{\beta T}^{\rm (soft)}\!=\!
{\rm Corr}(\beta_s,T_{\rm kin})$,
$\rho_{\beta T}^{\rm (mid)}\!=\!
{\rm Corr}(\beta_{\rm mid},T_q)$,
and
$\rho_{\rm sm}\!=\!
{\rm Corr}(\beta_s,\beta_{\rm mid})$.
Each draw defines one event, evaluated consistently for all sectors, and
two slightly decorrelated subevents are generated to mimic the experimental
$\eta$-subevent method and suppress self-correlations. 
The two subevents share the same underlying event parameters but include
independent statistical noise, ensuring that the reference quantity used
in $v_0(p_T)$ is decorrelated from the particle of interest.

Quenching-like fluctuations are introduced through two deterministic
schemes that couple the hard yield to the soft background.
In the amplitude modulation mode, the hard component scales as
$H'(p_T)=A\,H(p_T)$ with
$A=1+\sigma_A(w_\beta\Delta\beta/\beta_0+w_T\Delta T/T_0)$,
representing global yield suppression or enhancement correlated with the
soft flow.
In the slope modulation mode, the power-law index fluctuates as
$n_{\rm eff}=n_0[1+\sigma_n(w_\beta\Delta\beta/\beta_0+w_T\Delta T/T_0)]$,
yielding
$H'(p_T)\propto(1+p_T/p_0)^{-n_{\rm eff}}$
after normalization. Both prescriptions provide controlled soft–hard linkages: amplitude modulation generates an anti–correlated change in the hard yield relative to the soft flow, while slope modulation produces a partially correlated tilt of the hard tail without altering its normalization.
The default analysis adopts the slope–modulation configuration, where
the hard index responds primarily to the soft velocity
($w_\beta\!=\!2.0$, $w_T\!=\!0$) with a fractional width
$\sigma_n\!=\!0.1$.
Together, these prescriptions enable tunable soft–hard correlations and
quantify how coherent, partially correlated, and anti-correlated modes
shape $v_0(p_T)$ and $V_0(p_{T1},p_{T2})$.

Event-by-event variations of the soft and mid-$p_T$ parameters are drawn
from a correlated Gaussian ensemble characterized by
$\sigma_{\beta_s}\!=\!0.012$, $\sigma_{T_{\rm kin}}\!=\!0.002$, and
$\rho_{\beta T}^{\rm (soft)}\!=\!-0.5$ for the soft sector, and
$\sigma_{\beta_{\rm mid}}\!=\!0.04$, $\sigma_{T_q}\!=\!0.03$, and
$\rho_{\beta T}^{\rm (mid)}\!=\!-0.55$ for the coalescence sector.
The inter-sector correlation between soft and mid flow velocities is set
to $\rho_{\rm sm}\!=\!0.5$.

Table~\ref{tab:params} lists representative baseline parameters for all
three sectors and their corresponding fluctuation widths and
correlations.
Values are chosen to reproduce generic LHC Pb--Pb spectra and
$v_0(p_T)$.
The resulting decomposition of the spectra is shown in
Fig.~\ref{fig:spectra}, which captures the measured slopes and
illustrates the successive dominance of soft, mid, and hard
contributions that underlie the fluctuation analysis in the main text.

\begin{table}[t]
\centering
\caption{Representative baseline and fluctuation parameters of the three-component model.}
\label{tab:params}
\vspace{0.3em}
\begin{tabular}{lccc}
\hline
\hline
Component & Parameter & Symbol & Value \\
\hline
Soft
 & Kinetic temperature [GeV] & $T_{\rm kin}$ & 0.094 \\
 & Surface velocity & $\beta_s$ & 0.897 \\
 & Profile exponent & $n$ & 0.739 \\
 & Dispersion of $\beta_s$ & $\sigma_{\beta_s}$ & 0.012 \\
 & Dispersion of $T_{\rm kin}$ [GeV] & $\sigma_{T_{\rm kin}}$ & 0.002 \\
 & Intra-sector corr. & $\rho_{\beta T}^{\rm (soft)}$ & $-0.5$ \\
\hline
Mid
 & Quark slope [GeV] & $T_{q}$ & 0.36 \\
 & Flow exponent & $\eta$ & 1.8 \\
 & Reference velocity & $\beta_{\rm ref}$ & 0.65 \\
 & Tail scale [GeV] & $p_s$ & 4.7 \\
 & Tail exponent & $\delta$ & 1.2 \\
 & Dispersion of $\beta_{\rm mid}$ & $\sigma_{\beta_{\rm mid}}$ & 0.04 \\
 & Dispersion of $T_q$ [GeV] & $\sigma_{T_q}$ & 0.03 \\
 & Intra-sector corr. & $\rho_{\beta T}^{\rm (mid)}$ & $-0.55$ \\
 & Inter-sector corr. & $\rho_{\rm sm}$ & 0.5 \\
\hline
Hard
 & Scale [GeV] & $p_0$ & 2.0 \\
 & Power index & $n_0$ & 5.8 \\
 & Slope fluctuation width & $\sigma_n$ & 0.1 \\
 & Velocity coupling weight & $w_\beta$ & 2.0 \\
 & Temperature coupling weight & $w_T$ & 0.0 \\
\hline
\hline
\end{tabular}
\end{table}

\end{document}